\documentclass[11pt]{article}
\usepackage{CJKutf8}
\usepackage{amsmath,amssymb,color,graphics,epsfig,tikz}

\DeclareMathOperator{\sech}{sech}
\DeclareMathOperator{\csch}{csch}
\usepackage{amsfonts}

\newcommand{\be}{\begin{equation}}
\newcommand{\ee}{\end{equation}}
\newcommand{\bea}{\setlength\arraycolsep{2pt} \begin{eqnarray}}
\newcommand{\eea}{\end{eqnarray}}
\newcommand{\nn}{\nonumber}

\def\fft#1#2{{\frac{#1}{#2}}}

\def\0{{\sst{(0)}}}
\def\1{{\sst{(1)}}}
\def\2{{\sst{(2)}}}
\def\3{{\sst{(3)}}}
\def\4{{\sst{(4)}}}
\def\5{{\sst{(5)}}}
\def\6{{\sst{(6)}}}
\def\7{{\sst{(7)}}}
\def\8{{\sst{(8)}}}
\def\sst#1{{\scriptscriptstyle #1}}

\begin{document}
\begin{CJK}{UTF8}{gbsn}

\begin{flushright}
\end{flushright}

\vspace{25pt}
\begin{center}
{\large {\bf A new class of exactly-solvable potentials by means of the hypergeometric equation}}

\vspace{10pt}
Wei Yang

\vspace{10pt}

{\it College of Science, Guilin University of Technology, Guilin, Guangxi 541004, China}

\vspace{40pt}

\underline{ABSTRACT}
\end{center}

We obtained a new class of exactly-solvable potentials by means of the hypergeometric equation for Schr\"{o}dinger equation, which different from the exactly-solvable potentials introduced by Bose and Natanzon.  Using the new class of solvable potentials, we can obtain the corresponding complex PT-invariant potentials. This method can also apply to the other Fuchs equations.

\vfill {\footnotesize Emails: weiyang@glut.edu.cn}
\thispagestyle{empty}

\pagebreak



\newpage

\section{Introduction}
\label{Sec.introduce}

The exact solutions to the Schr\"{o}dinger equation play crucial roles in quantum physics. It is well-known that there are several potentials can be exactly solved, as examples one can site the
harmonic oscillator, the Coulomb, the Morse \cite{Morse}, P\"{o}schll-Teller \cite{Teller}, Eckart \cite{Eckart:1930zza} potentials and so on. The reason why these potentials are exact solvable is that the Schr\"{o}dinger equation in these potentials can be transformed by into either the hypergeometric or the confluent hypergeometric equation.
Here we focus on hypergeometric equation to construct a new class of the solvable potentials within the framework of the non-relativistic Schr\"{o}dinger equation. Similar works have been carried out \cite{Bose,Natanzon,Ishkhanyan,Dong}. 
Using the new class of solvable potentials that we have constructed, we can easily obtain complex PT-invariant potentials, these potentials is a hot issue of recent study\cite{Ahmed:2001gz,Kumari,Hasan:2017fje,Yadav:2015fia}.

This work is organized as follows. In Section \ref{Sec.formalism}, We present the basic method of our argument. In Section \ref{Sec.Hypergeometric} as an illustration we take the hypergeometric equation to
construct the soluble potentials. In Section \ref{Sec.conclusion} we will give some conclusions for this paper.

\section{Basic methods}
\label{Sec.formalism}
As we know, the stationary Schr\"{o}dinger equation for a particle of energy $E$ in a potential $V(r)$ has the form
\begin{align} 
\psi''(r)+(E-V(r))\psi(r)=0\,.
\label{eq:0}
\end{align} 
Through choosing a transformation of the variable $z=z(r)$, the equation become
\begin{align}
\psi_{zz}+\fft{\rho_z}{\rho}\psi_z+\fft{E-V}{\rho^2}\psi=0\,.
\end{align}
Where $\rho=\fft{dz}{dr}$, and apply the transformation $\psi=f(z)u(z)$. we can obtain the following differential equation
\begin{align}
u_{zz}+(2\fft{f_z}{f}+\fft{\rho_z}{\rho})u_z+(\fft{f_{zz}}{f}+\fft{f_z}{f}\fft{\rho_z}{\rho}+\fft{E-V}{\rho^2})\psi=0\,.
\end{align}
Compared with our target equation
\begin{align}
u_{zz}+g(z)u_z+h(z)u=0\,.
\label{eq:1}
\end{align}
we have
\begin{align}
&g(z)=2\fft{f_z}{f}+\fft{\rho_z}{\rho}\,\label{eq:2}\\
&h(z)=\fft{f_{zz}}{f}+\fft{f_z}{f}\fft{\rho_z}{\rho}+\fft{E-V}{\rho^2}\,.\label{eq:3}
\end{align}
Integrating the equation (\ref{eq:2}) allows us to obtain
\begin{align}
&f(z)=\sqrt{c_1}\rho^{-1/2}e^{\int g(z)dz/2}\,.
\end{align}
Substitution of this equation into (\ref{eq:3}) yields
\begin{align}
E-V=\rho^{2}(h-\fft{g_z}{2}-\fft{g^2}{4})+\fft{1}{2}\{z,r\}\,.
\label{eq:8}
\end{align}
Where the Schwarzian derivative given as
\begin{align}
\{z,r\}=\fft{z'''(r)}{z'(r)}-\fft{3}{2}(\fft{z''(r)}{z'(r)})^2=\rho\rho_{zz}-\fft12\rho_z^2\,.
\end{align}
Therefore, on the basis of the solvable differential equation (\ref{eq:1}),  we are able to construction of the solvable potentials for the original Schr\"{o}dinger equation. This problem will become how to find that new variable $z(r)$ and function $f(z)$ satisfy equation (\ref{eq:2}) (\ref{eq:3}).

\section{Hypergeometric equation}
\label{Sec.Hypergeometric}
In this section, we application the methods to hypergeometric equation
\begin{align}
z(1-z)u''(z)+[\gamma-(\alpha+\beta+1)z]u'(z)-\alpha\beta u(z)=0\,.
\end{align}
Obviously, in this case $g(z)=\fft{\gamma-(\alpha+\beta+1)z}{z(1-z)}$ and $h(z)=-\fft{\alpha\beta}{z(1-z)}$.

In order to determine the form of $f$, we follow the Ishkhanyan et al~\cite{Ishkhanyan} who discussed the reduction of the Schr\"{o}dinger equation to a rather large class of target equations by a transformation of  $\rho(z)$, and $\rho(z)$ is a polynomial, which all the roots should necessarily coincide with the singular points of the target equation to which the Schrödinger equation is reduced. 
Then, the coordinate transformation of $f$ need the form
$f=z^p(1-z)^q$, so the equation (\ref{eq:2}) have solution
\begin{align}
\rho=c_1z^{\gamma-2p}(1-z)^{\alpha+\beta+1-\gamma-2q}\,.
\label{eq:4}
\end{align}
Substitution of this equation (\ref{eq:4}) and $h(z)=-\fft{\alpha\beta}{z(1-z)}$ into (\ref{eq:3}), we can easily figure out that the Schr\"{o}dinger equation have solvable potentials only if $a=\gamma-2p$ and $b=\alpha+\beta+1-\gamma-2q$ satisfy some of these discrete values $a=0,1,2$, $b=0,1,2$, and in these cases the functional form of $z(r)$ and $V(r)$ are likely to be simple.

\textbf{Case:1} consider $a=0,b=1$, so $\rho=\fft{dz}{dr}=c_1(1-z)$, therefore
\begin{align}
&z=1+c_2e^{-c_1r}\,.  
\end{align}
From equation (\ref{eq:8}), we can get 
\begin{align}
&E-V=C-A\fft{1}{(c_2+e^{c_1r})^2}-B\fft{1}{c_2+e^{c_1r}}\,.  
\end{align}
Where the coefficients are given
\begin{align}
&A=\fft{c_1^2c_2^2}{4}\gamma(\gamma-2)\nn\\
&B=\fft{c_1^2c_2}{2}(\alpha\gamma+\beta\gamma+\gamma-\gamma^2-2\alpha\beta)\nn\\
&C=-\fft{c_1^2}{4}(\alpha+\beta-\gamma)^2
\,.  
\end{align}

\textbf{Case:2}  consider $a=1/2,b=1/2$ , $\rho=c_1z^{1/2}(1-z)^{1/2}$, therefore
\begin{align}
&z=1-\sin(1/2(c_1r+c_2))^2\,.  
\end{align}
So equation (\ref{eq:8}) tells us that
\begin{align}
&E-V=C-A\csc(c_1r+c_2)^2-B\cot(c_1r+c_2)\csc(c_1r+c_2)\,.  
\end{align}
Where the coefficients are given
\begin{align}
&A=\fft{c_1^2}{4}(2(\alpha+\beta)^2+1)-c_1^2(\alpha+\beta-\gamma+1)\gamma\nn\\
&B=\fft{c_1^2}{2}(\alpha+\beta-1)(\alpha+\beta+1-2\gamma)\nn\\
&C=\fft{c_1^2}{4}(\alpha-\beta)^2
\,.  
\end{align}

\textbf{Case:3} consider $a=1/2,b=1$ , $\rho=c_1z^{1/2}(1-z)$ , therefore
\begin{align}
&z=\tanh^2(\fft{c_1r-c_2}{2})\,.  
\end{align}
So we have
\begin{align}
&E-V=C-A\sech^2[(c_1r-c_2)/2]-B\csch^2[(c_1r-c_2)/2]\,.  
\end{align}
Where the coefficients are given
\begin{align}
&A=-\fft{c_1^2}{16}(4(\alpha-\beta)^2-1)\nn\\
&B=\fft{c_1^2}{16}(4\gamma^2-8\gamma+3)\nn\\
&C=-\fft{c_1^2}{4}(\alpha+\beta-\gamma)^2
\,.  
\end{align}

\textbf{Case:4} consider $a=1,b=0$, $\rho=c_1z$, therefore 
\begin{align}
&z=c_2e^{c_1r}\,.  
\end{align}
we have
\begin{align}
&E-V=C-A\fft{1}{(c_2e^{c_1r}-1)^2}-B\fft{1}{c_2e^{c_1r}-1}\,.  
\end{align}
Where the coefficients are given
\begin{align}
&A=-\fft{c_1^2}{4}((\alpha+\beta-\gamma)^2+1)\nn\\
&B=\fft{c_1^2}{2}(\alpha^2+\beta^2+\gamma(1-\alpha-\beta)-1)\nn\\
&C=-\fft{c_1^2}{4}(\alpha-\beta)^2
\,.  
\end{align}

\textbf{Case:5} consider $a=1,b=1/2$ , $\rho=c_1z(1-z)^{1/2}$ , therefore 
\begin{align}
&z=\sech^2(\fft{c_1r+c_2}{2})\,.
\end{align}
we have
\begin{align}
&E-V=C-A\csch^2(c_1r+c_2)-B\coth(c_1r+c_2)\csch(c_1r+c_2)\,.  
\end{align}
Where the coefficients are given
\begin{align}
&A=\fft{c_1^2}{4}[4(\alpha^2+\beta^2)-4(\alpha+\beta)+2\gamma-1]\nn\\
&B=\fft{c_1^2}{2}(2\alpha-\gamma)(2\beta-\gamma)\nn\\
&C=-\fft{c_1^2}{4}(\gamma-1)^2
\,.  
\end{align}

\textbf{Case:6} consider $a=1,b=1$ , $\rho=c_1z(1-z)$, therefore 
\begin{align}
&z=1-\fft{c_2}{e^{c_1r}+c_2}\,.  
\end{align}
we have
\begin{align}
&E-V=C-A\fft{e^{2c_1r}}{4(e^{c_1r}+e^{c_2})^2}-B\fft{e^{c_1r}}{4(e^{c_1r}+e^{c_2})}\,.  
\end{align}
Where the coefficients are given
\begin{align}
&A=\fft{c_1^2}{4}[(\alpha-\beta)^2-1]\nn\\
&B=\fft{c_1^2}{2}(2\alpha\beta+\gamma-\alpha\gamma-\beta\gamma)\nn\\
&C=-\fft{c_1^2}{4}(\gamma-1)^2
\,.  
\end{align}
So far we have discussed six cases, which represent six possible types of potential energy. The other three possible cases
$a=0,b=0$; $a=0,b=1/2$ and $a=1/2,b=0$ cannot be considered as solvable potential due to the lack of an energy terms.

Further, by applying the relationship between the hypergeometric function and the Riemannian $P$-function and its transformation formula, we can obtain other solutions of different forms.

It is worth noting that in case 2 if we set $c_1=i,c_2=\pi/2$, then 
\begin{align}
&z=(1-i\sinh r)/2\,.  
\end{align}
And 
\begin{align}
&E-V=C-A\sech^2 r+iB\sech r\tanh r\nn\\
&A=-\fft{1}{4}(2(\alpha+\beta)^2+1)+(\alpha+\beta-\gamma+1)\gamma\nn\\
&B=-\fft{1}{2}(\alpha+\beta-1)(\alpha+\beta+1-2\gamma)\nn\\
&C=-\fft{1}{4}(\alpha-\beta)^2
\,.  
\end{align}
This is nothing but the complex PT-invariant potential energy discussed by Ahmed et al~\cite{Ahmed:2001gz,Kumari}.
Similarly, we can construct others solvable PT-invariant  potentials.  Set $c_1$ is pure imaginary number and $c_2$ is real number in all sex cases, we will find these potentials are all complex PT-invariant potentials. 

If set $g(z)=0$ in the equation (\ref{eq:2}), then we can solve $f(z)=\sqrt{c_1}\rho^{-1/2}$, and also assume that $f(z)$ have the form
$f=z^p(1-z)^q$, then we will arrive at the Bose solvable potentials~\cite{Bose}. These potentials have been studied generally~\cite{Ishkhanyan,Natanzon,Milson:1997cp,Morales}. 
In a similar way, we can also construct the corresponding complex PT-invariant potentials.
\section{Conclusion}
\label{Sec.conclusion}

In this work, we construct a new class of solvable potentials by means of the hypergeometric equation, where the Schr\"{o}dinger equation is rewritten as a hypergeometric equation through taking a similarity transformation. we consider $g(r)\neq 0$, and suppose $f=z^p(1-z)^q$, such that $p,q$ dependent $\alpha,\beta,\gamma$, this method provides us a new perspective to construct solvable  potentials for Schr\"{o}dinger equation.
Our method unlike previously let $g(r)=0$ or used the invariant identity of the target equation, this method need $f(z)=\sqrt{c_1}\rho^{-1/2}$, cause $f(z)$ to be a fixed function.
In a similar way, computing the other Fuchs equations with our method, such as the Heun equation, can also construct corresponding solvable potentials\cite{Filipuk}.

\section*{Acknowledgement}
This work was supported by the Guangxi Scientific Programm Foundation under grant No. 2020AC20014, the Scientific Research Foundation of Guilin University of Technology under grant No. GUTQDJJ2019206.

\end{CJK}
\end{document}